\documentclass[preprint,amsmath,amssymb]{revtex4}

\usepackage{graphicx}
\usepackage{dcolumn}
\usepackage{bm}

\begin{document}

\title{Stability measurements of microwave frequency synthesis with liquid helium cooled cryogenic sapphire oscillators}

\author{J.G.~Hartnett,$^1$ D.L.~Creedon,$^1$ D.~Chambon$^2$ and G.~Santarelli$^2$}
 \email{john@physics.uwa.edu.au}
\affiliation{$^1$School of Physics, University of Western Australia 35 Stirling Hwy Crawley 6009 W.A. Australia\\ $^2$Laboratoire National de M$\acute{e}$trologie et d$\acute{}$Essais Syst$\grave{e}$mes de R$\acute{e}$f$\acute{e}$rence Temps-Espace (LNE-SYRTE),
Observatoire de Paris, 61 Avenue de l$\acute{}$Observatoire, 75014 Paris, France}

\date{\today}

\begin{abstract}
We report on  the evaluation of microwave frequency synthesis using two cryogenic sapphire oscillators developed at the University of Western Australia. A down converter is used to make comparisons between microwave clocks at different frequencies, where the synthesized signal has a stability not significantly different from the reference oscillator. By combining the CSO with a H-maser, a reference source of arbitrary frequency at X-band can be synthesized with a fractional frequency stability of sub-$4 \times 10^{-15}$ for integration times between 1 s and $10^4$ s.
\end{abstract}

\maketitle

\section{Introduction}
Ultra stable oscillators are used as part of many measurement systems in time and frequency metrology \cite{Tobar2006}. Cryogenic sapphire oscillators (CSO) have been developed at the University of Western Australia (UWA) over the past two decades. They are state-of-the-art \cite{Chang2000} with fractional frequency stability for a single CSO less than or equal to $10^{-15}$ for integration times between 1 and 1000 s \cite{Hartnett2006, Locke2008}. CSOs are now used in research labs around the world for applications including a flywheel microwave reference for atomic fountain and cold atom clocks \cite{Tobar2006, Fisk1995, Santarelli1999}, the RF reference for ultra-stable frequency combs \cite{Watabe2007} and as instruments to test fundamental physics such as in modern Michelson-Morley and Kennedy-Thorndike experiments \cite{Wolf2003, Lemonde1999, Stanwix2005, Stanwix2006}. 

Atomic fountain clocks have been operated now for more that 10 years as a frequency standard \cite{Clairon1995}, and are soon expected to reach an accuracy of $10^{-16}$ or better. Approaching this level of accuracy all systematic effects have to be evaluated, and the stability requirement driving the atomic fountain is quite stringent. Currently, to meet this, CSOs are used with a number of fountain or cold atom clocks in various labs. Different frequency synthesis designs have been implemented and reported in the literature \cite{Rovera1996, Gupta2000, Chambon2005, Chambon2007}. Synthesis chains are needed to produce the desired microwave frequency. In this paper we report on the development of an additional liquid helium cooled CSO at UWA, developed for the purpose of providing a better VLBI radio-astronomy reference oscillator than a H-maser, and the frequency synthesizers used to make comparisons between microwave clocks at different frequencies, where the synthesized signal has a stability not significantly different from that of the reference oscillator.  This is of interest to radio-astronomers to see whether synthesis methods can deliver extremely high quality oscillator stability at arbitrary microwave frequencies by scaling up from 100 MHz. The CSO will be shipped to M.I.T. Haystack Observatory in mid 2010 where it will be used with a purpose build synthesizer to generate a stable 10 MHz reference.

\section{Cryogenic sapphire oscillators}
\begin{figure}[!t]
\centering
\includegraphics[width=3.5in]{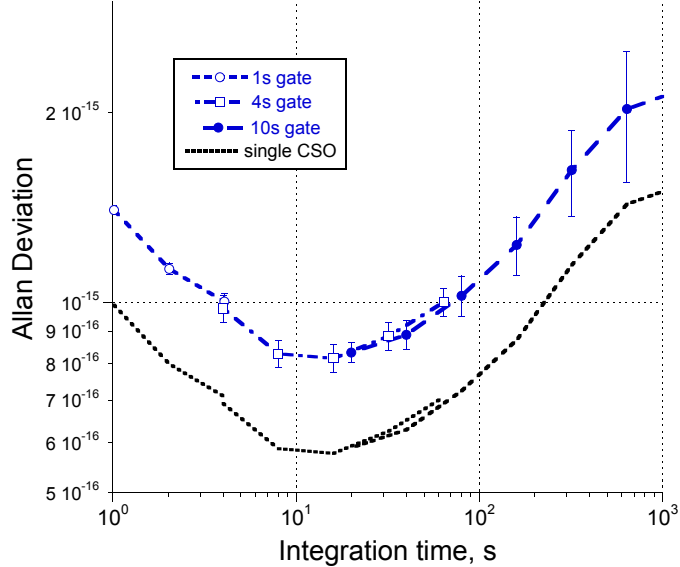}
\caption{Allan Deviation of the CSOs determined from the fractional frequency of the 332 kHz beat note between the two nominally identical oscillators (top curve) and the inferred stability of a single oscillator (bottom curve) assuming they contribute equally.}
\label{fig_1}
\end{figure}

A Pound stabilized, power stabilized CSO (named here CSO2) was built as a nominally identical pair to the existing CSO1 in the lab. It operates on the same $WGH_{16,0,0}$ mode as the latter. Its frequency is 11.200053841 GHz, its loaded Q-factor is $4.4 \times 10^8$ with a primary coupling of $0.61$ at its frequency-temperature turning point of 6.6719 K. In contrast CSO1 has a frequency of 11.200386531 GHz, a loaded Q-factor of $7 \times 10^8$ with a primary coupling of unity at its  frequency-temperature turning point of 7.2319 K. The only other difference is that output power of the microwave amplifier in the sustaining loop is about 6 dB greater hence nearly 14 dBm of output power is derived from CSO2, as compared with only 2 dBm from CSO1. Since the loaded Q-factor is critical to the electronic noise floor of the Pound servo the primary coupling on CSO2 was not further increased toward critical. Nevertheless, the final oscillator stability as shown in Fig. \ref{fig_1} is excellent. The upper dashed curve in Fig. \ref{fig_1} is the Allan Deviation calculated from the 332 kHz beat note between the two oscillators, filtered then counted on an Agilent 53132A $\Lambda$-type counter. The lower dotted curve is the inferred single CSO stability assuming both contribute equally \cite{Dawkins}.  However it is believed, but yet to be confirmed by a 3 cornered hat measurement with 3 CSOs, that CSO2 has poorer stability in the short term due to reduced resonator Q-factor.

In both CSOs about 1 mW of power is incident on the cryogenic resonator. This is optimal, to avoid saturating the Pound servo detector located on the reflection port of a circulator connected to the primary coupling probe in the cryogenic cavity (since CSO2 has non-critical coupling), to avoid saturating the power servo detector located on a 10 dB directional coupler in the input transmission line in the cryogenic environment outside the cavity, to minimize the frequency instability resulting from radiation pressure on the resonator and to minimize the power dependence of spurious AM to PM conversion due to the frequency modulator having an insertion loss dependence on bias voltage. This result underlines what can be achieved with a resonator that was less than optimum in terms of the resonant mode quality factor at cryogenic temperatures. Critical coupling and a loaded Q-factor of $7 \times 10^8$ is not essential.

\section{100 MHz synthesizer}
\begin{figure}[!t]
\centering
\includegraphics[width=3.5 in]{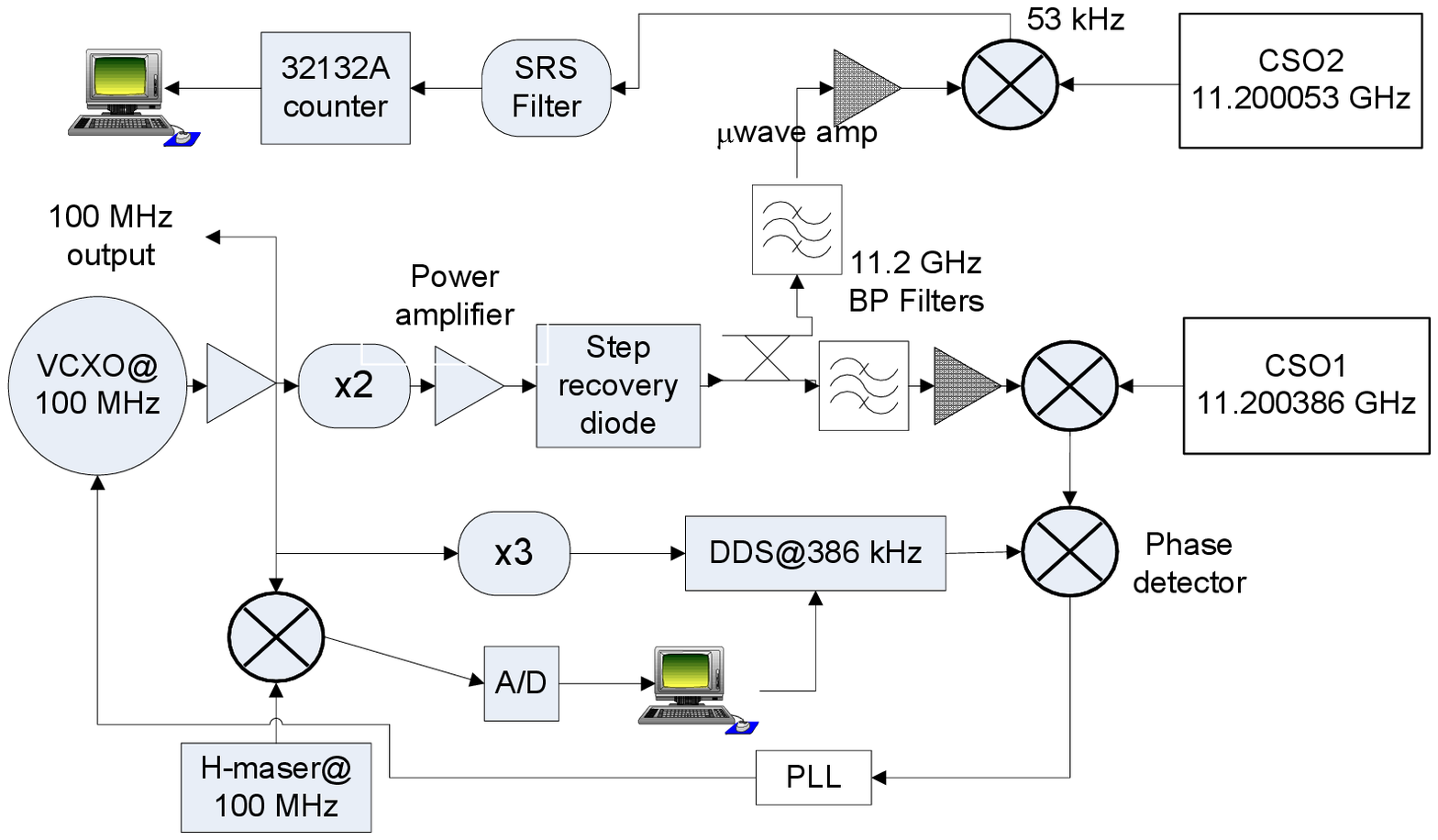}
\caption{Block diagram of the down conversion synthesis to 100 MHz from 11.200386 GHz of CSO1 and the up conversion to 11.200000 GHz, via the 56th harmonic of the step recovery diode comb used in the phase lock of the CSO1 to the VCXO. The result is compared with 11.200053 kHz of CSO2.}
\label{fig_2}
\end{figure}

Two nominally identical synthesizers were built to down-convert an X-band signal to 100 MHz. Fig. \ref{fig_2} shows the block diagram of the component layout that was used. The output signal of a low phase noise 100 MHz VCXO quartz is tightly phase locked, with an analog circuit, to the signal derived from a cryogenic sapphire oscillator (CSO1 in Fig. \ref{fig_2}). This is achieved by up-converting the 100 MHz signal to 11.200386 GHz by picking off the appropriate the step recovery diode (SRD) comb member corresponding to 11.2 GHz and adding the difference. The  386 kHz difference in this case is generated using a DDS module driven by a 300 MHz clock frequency derived from the 100 MHz of the VCXO. In addition, over longer time scales, a digital phase lock is enabled causing the 100 MHz VCXO to track the 100 MHz  from a H-maser.  We call this configuration I. Note the SRD was driven hard with about 1 W of power from the 200 MHz signal. It is expected that this should limit the fractional frequency stability of the synthesis at short to medium integration times.

\begin{figure}[!t]
\centering
\includegraphics[width=3.5in]{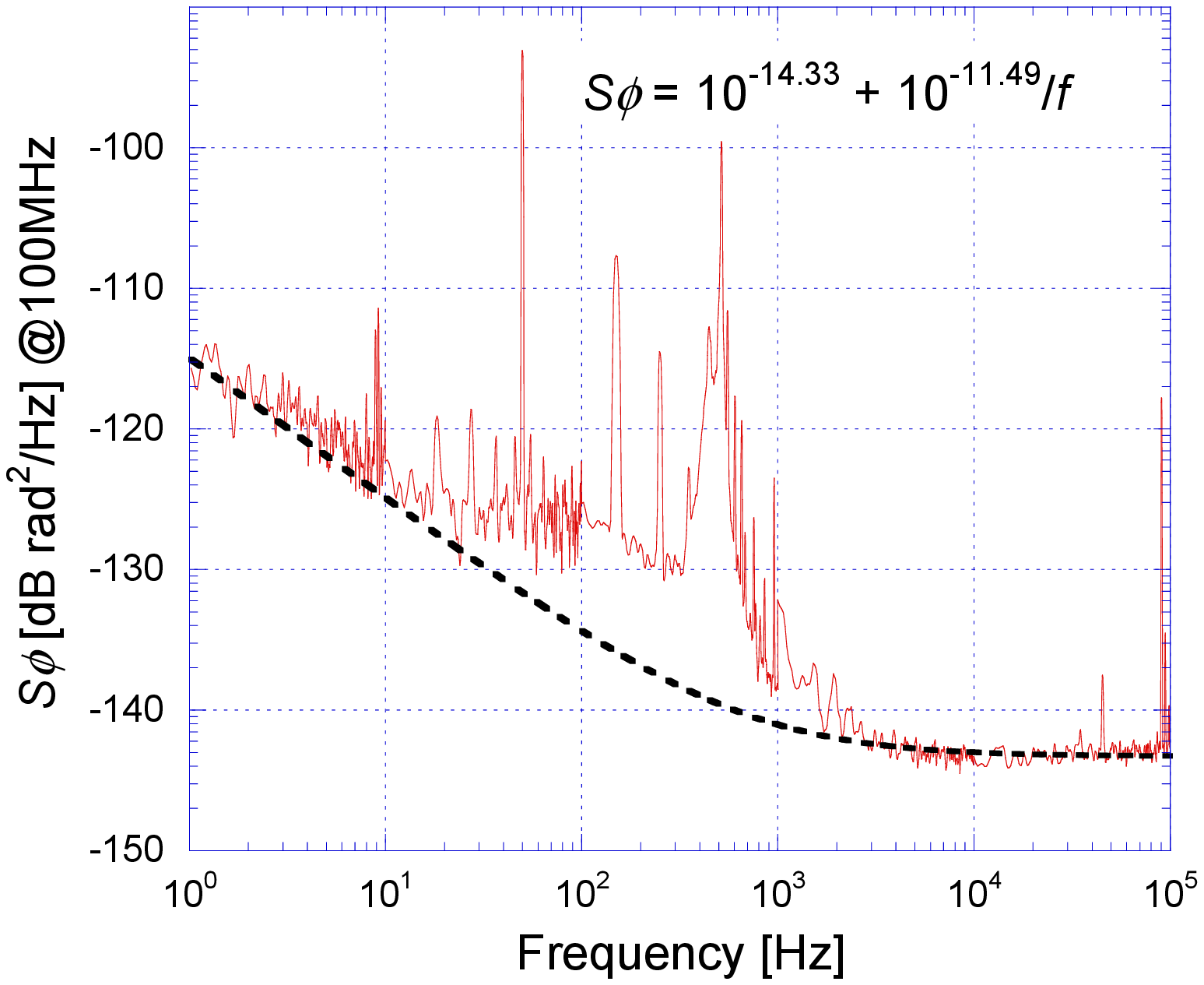}
\caption{Residual phase noise spectral density of two nominally identical down converters at 100 MHz (red/solid curve) and the best fit power laws (black/broken curve).}
\label{fig_3}
\end{figure}

\begin{figure}[!t]
\centering
\includegraphics[width=3.5in]{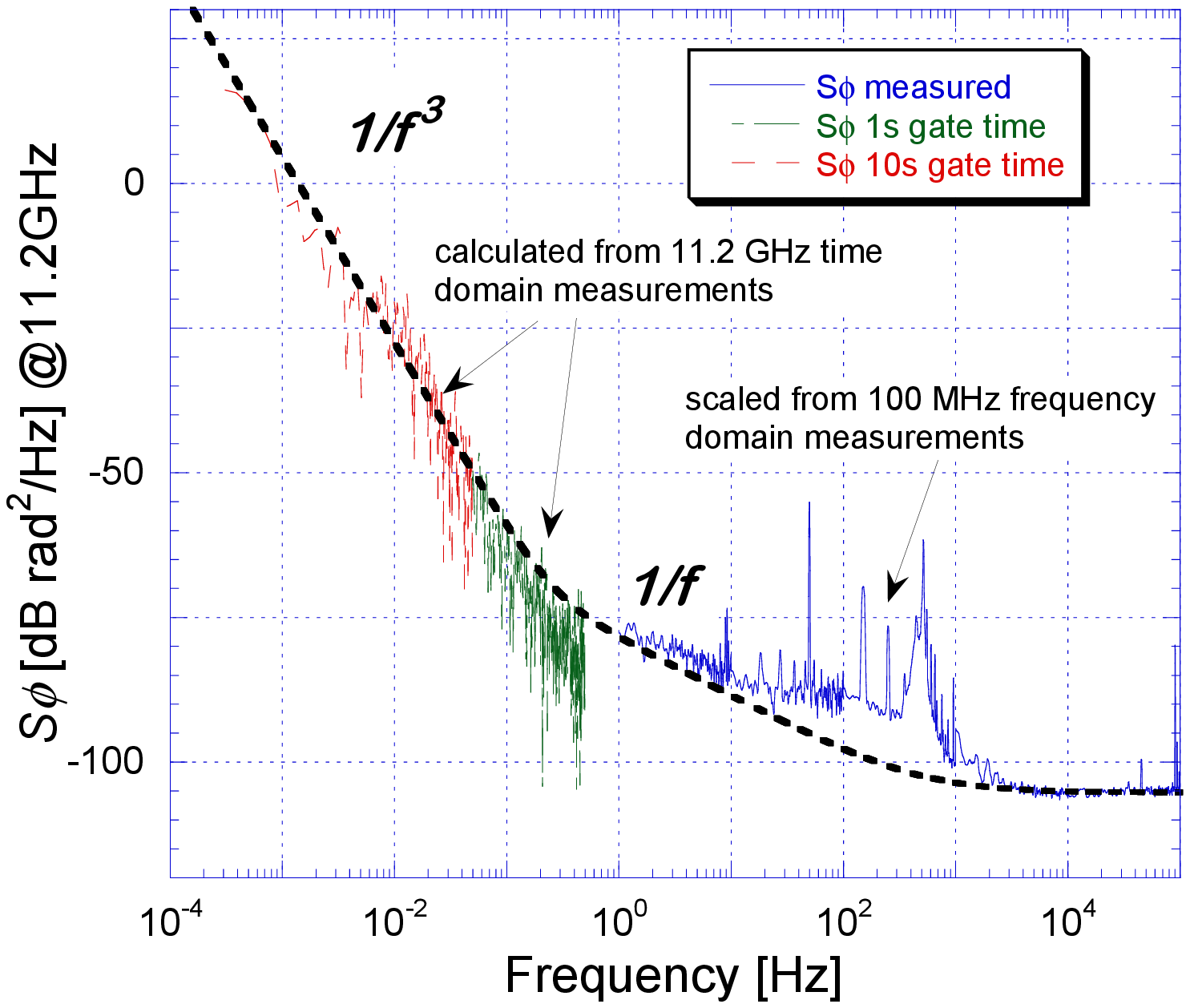}
\caption{Phase noise $S_{\phi}$ of the synthesizer at 11.200 GHz determined from a) scaled down converter residual phase noise measurements (blue curve) and b) $S_y$ calculated from spectral density of time domain measurements of the beat between the synthesizer at 11.200 GHz and the other CSO using 1s and 10s gate times on the frequency counter (green and red curves).}
\label{fig_4}
\end{figure}

The residual phase noise of two nominally identical down converters was measured at 100 MHz, where the pump signal was derived from the same low noise source, and it is shown in Fig. \ref{fig_3}. After reducing the latter by 3 dB for a single down converter and scaling to 11.2 GHz we derive the phase noise for the synthesis at the microwave frequency as shown in Fig. \ref{fig_4} at Fourier frequencies above 1 Hz. The synthesizer phase noise at 11.2 GHz is about -80 dBc at 1 Hz offset from the carrier.

\begin{figure}[!t]
\centering
\includegraphics[width=3.5in]{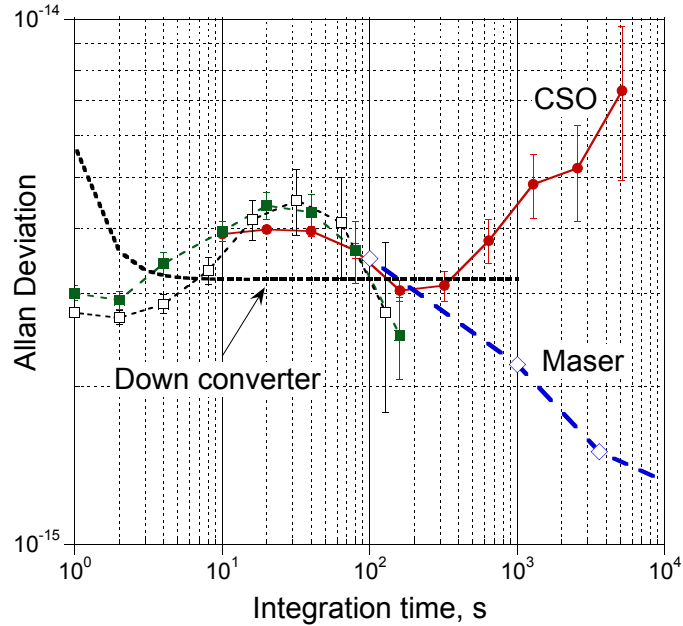}
\caption{Allan Deviation of the fractional frequency fluctuations of the synthesizer at 11.200 GHz using configuration I (using CSO1, red solid circles at 10s gate time and green solid squares at 1s, 2s, 4s and 10s gate times) and configuration II (using CSO2,  black open square at 1s gate time).  The estimated stability contributions from the down converter (black broken curve) and from the H-maser (blue diamonds) are also shown for comparison.}
\label{fig_6}
\end{figure}

Using configuration I, the 56th harmonic of the SRD was then selected and band pass filtered at 11.2 GHz and directly compared with  CSO2, by counting the 53 kHz beat on an Agilent 53132A $\Lambda$-counter. The Allan deviation of the resulting beat was computed and is shown in Fig. \ref{fig_6}. The data from 1 s and 10 s gate times were converted to power spectral density (PSD) and the equivalent phase noise calculated. This is shown in Fig. \ref{fig_4} at Fourier frequencies below 1 Hz. The agreement with the residual phase noise measurement of the down converters is quite good indicating that the CSOs contribute very little to the phase noise of the synthesis or the measurement.

\begin{figure}[!t]
\centering
\includegraphics[width=3.5in]{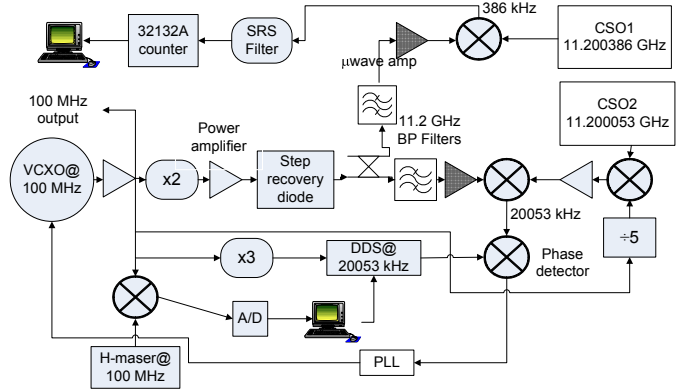}
\caption{Block diagram of the down conversion synthesis to 100 MHz from 11.200053 GHz of CSO2 and the up conversion to 11.200000 GHz, via the 56th harmonic of the step recovery diode comb used in the phase lock of the CSO2 to the VCXO. The result is compared with 11.200386 kHz of CSO1.}
\label{fig_5}
\end{figure}

When attempting to run the synthesizer from the output of CSO2, which differs from 11.200 GHz by only 53 kHz, the digital PLL would not work since the DDS units are AC coupled. To solve this, configuration II was implemented where a 20 MHz signal, derived by dividing (by 5) the 100 MHz of the down converter output with a Hittite low noise divider, was mixed with the output of CSO2, since it had plenty of available power.  This was filtered and injected into the down converter of configuration I. See Fig. \ref{fig_5}. 

Similarly the 56th harmonic of the SRD was used to compare again with the other CSO (CSO1 in this case) at 386 kHz. The Allan deviation of the beat note is shown in Fig. \ref{fig_6}. The central hump for integration times $10<\tau<100$ s in Fig. \ref{fig_6} is due to too much proportional gain in the digital PLL hence locking to the hydrogen maser signal too tightly. The lock to the H-maser is apparent and seen where the maser stability (shown by diamonds in Fig. \ref{fig_6}) is comparable with the measured stability. The long term dependence results from the reference CSO1 used to make the comparison. The stability measurements of both configurations, I and II, are shown in Fig. \ref{fig_6} and compared to the stability derived from a fit to the phase noise of Fig.  \ref{fig_4}. The disagreement for integration times $\tau<4$ s is explained below.

\begin{figure}[!t]
\centering
\includegraphics[width=3.5in]{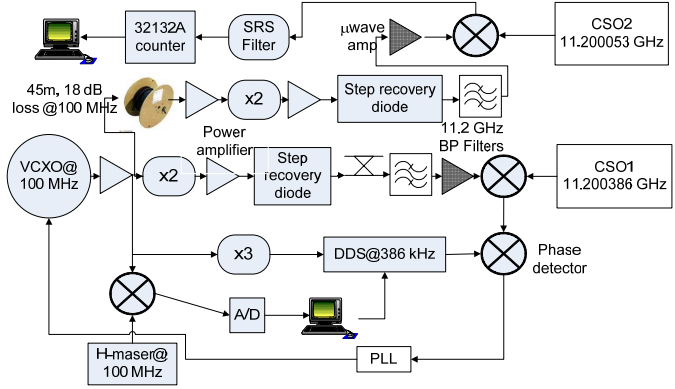}
\caption{Block diagram of the down conversion synthesis to 100 MHz from 11.200386 GHz of CSO1 and the up conversion to 11.200000 GHz, via 45 m of LMR-195 coax and the 56th harmonic of a second step recovery diode comb. The result is compared with 11.200053 kHz of CSO2.}
\label{fig_7}
\end{figure}

\begin{figure}[!t]
\centering
\includegraphics[width=3.5in]{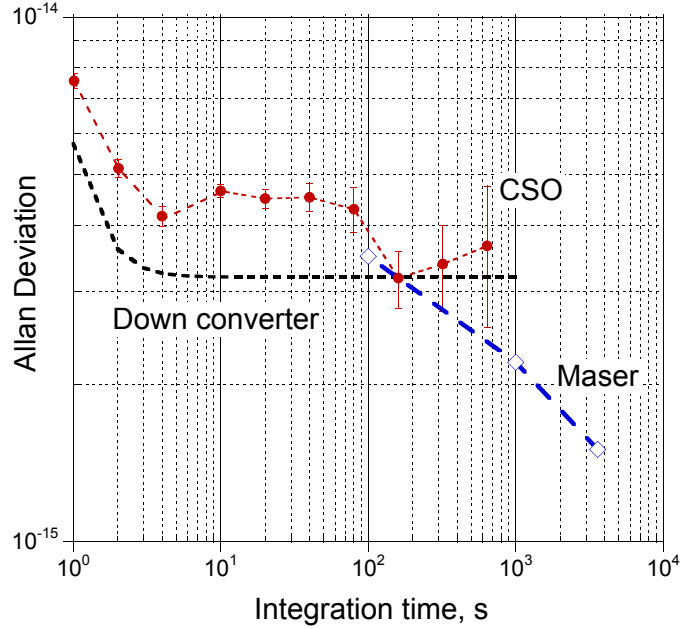}
\caption{Allan Deviation of the fractional frequency fluctuations of the synthesizer at 11.200 GHz using configuration III (red solid circles using CSO2 at 1s, 2s, 4s and 10s gate times).  The estimated stability contributions from the down converter (black broken curve) and from the H-maser (blue diamonds) are also shown for comparison.}
\label{fig_8}
\end{figure}

Finally, we wanted to evaluate the effect of sending the 100 MHz from the down converter over 45 m of LMR-195 coaxial cable. This is configuration III, shown in Fig. \ref{fig_7}. The cable introduced 18 dB of loss and thus it was required to introduce amplification sufficient to drive the doubler, the 200 MHz signal from which is amplified to 1 W to drive an additional SRD. Using again the 56th harmonic, which is filtered and amplified, an 11.2 GHz signal synthesized from CSO1 via the down converter is compared with that of CSO2 with a beat note of 53 kHz, which is counted and the result is shown in Fig. \ref{fig_8}. 

In this case also the proportional gain on the digital PLL was too high, locking too soon to the maser signal. A hump is apparent between integration times $10<\tau<100$ s in Fig. \ref{fig_6}. Nevertheless it is clear that the long term stability is due to the reference CSO, CSO2 in this case. For $\tau<2$ s the Allan Deviation has a $\tau^{-3/2}$ dependence and $2<\tau<4$ s it has a $\tau^{-1/2}$ dependence, from white phase and flicker phase of the amplifiers introduced in configuration III. The calculated Allan Deviation of the down converter stage is also shown in Fig. \ref{fig_8}, with short term $\tau$-dependence very similar to this. But the short term $\tau$-dependence of the down converter comes from a noisy component in one of the analog PLL mixer/amplifier boxes used in the two nominally identical units. This was discovered only after the phase noise measurements (as shown in Fig. \ref{fig_3}) were made. Hence we expect to see it in the reconstructed Allan deviation data, where the phase noise data from the frequency domain was used, but not in the results derived from the time series data, where that noisy PLL box had already been replaced with a low noise unit.

\section{Conclusion}
The fractional frequency stability of two cryogenic oscillators (one  constructed with a resonator loaded Q-factor of $4.4 \times 10^8$ and a primary coupling of only $0.61$) has been determined. The results are first class and one of the CSO is intended for use as a frequency reference for extremely high frequency VLBI radioastronomy. A frequency down converter was constructed and its stability evaluated using various configurations to make the comparisons near the microwave frequency (11.2 GHz) of the cryogenic oscillators. The results indicate that by combining the CSO with a H-maser, a reference source of arbitrary frequency at X-band can be synthesized with a fractional frequency stability of sub-$4 \times 10^{-15}$  for integration times between 1 s and $10^4$ s.

\section*{Acknowledgment}
The authors would like to thank M.E Tobar and E.N. Ivanov for useful discussions.

\end{document}